\documentstyle[graphicx,aps,multicol]{revtex}
\draft

\begin{document}

\title{Observation of the scissors mode in the quasicontinuum}
\author{A. Schiller\footnote{Electronic address: Andreas.Schiller@fys.uio.no},
M.~Guttormsen, E.~Melby, J.~Rekstad, and S.~Siem}
\address{Department of Physics, University of Oslo, N-0316 Oslo, Norway}
\author{A.~Voinov}
\address{Frank Laboratory of Neutron Physics, Joint Institute of Nuclear
Research, 141980 Dubna, Moscow reg., Russia}
\maketitle

\begin{abstract}
The experimental resonance parameters of the pygmy resonance in rare earth 
nuclei are compared to global observables of the scissors-mode states. It is 
argued that the pygmy resonance in rare earth nuclei can be described in terms 
of orbital M1 strength observed in $(\gamma,\gamma')$ experiments. The pygmy
resonance is therefore interpreted as the scissors mode in the quasicontinuum.
\end{abstract}


\begin{multicols}{2}

\section{Introduction}

The phenomena 'pygmy resonance' (PY) and 'scissors mode' (SC) have been
discussed separately in the literature. In this work, we investigate for the
first time quantitatively, if those two experimentally observed phenomena have
the same physical origin.

The PY was introduced as a name for non-statistical features in quasicontinuous
$\gamma$-ray spectra at energies below the giant dipole resonance. It is e.g.\
broadly discussed in a review of radiative strength functions as early as in
1973 \cite{BE73}. These non-statistical features observed in different nuclear 
mass regions and at different energies have different physical origins. A PY in
rare earth nuclei at an energy around 3.5~MeV was first reported in the 
$\gamma$-ray spectrum of $^{170}$Tm following a neutron capture reaction 
\cite{JD79}. A systematic investigation of the PY parameters in deformed rare 
earth nuclei by radiative neutron capture has been carried out in \cite{IK86}, 
and very recently new investigations on the PY have appeared 
\cite{MI99,VG00,MG00}. An important application of the PY is the calculation of
astrophysical (n,$\gamma$) reaction rates, where the existence of a soft dipole
mode can greatly enhance the calculated reactions rates in e.g.\ the 
astrophysical r-process \cite{Go98}.

The SC was discovered in 1984 in inelastic electron scattering experiments on 
$^{156}$Gd \cite{BR84}. The bulk of information on the SC comes, however, from 
$(\gamma,\gamma')$ experiments, also called nuclear resonance fluorescence 
(NRF). These data are compiled in Ref.\ \cite{PB98}, which also gives 
systematics of average excitation energies of SC states and summed orbital M1 
strengths around 3~MeV\@.

It exists no satisfactory theoretical explanation for the PY, and it is an
interesting thought that the PY and the SC might emerge due to the same 
physical origin, i.e.\ if the PY can be described in terms of the observed 
orbital M1 strength from NRF experiments. In order to investigate this idea, we
will compare the energy, the spreading width, and the resonance strength of the
two phenomena.

\section{Scissors mode}
\label{sec:scissors}

In NRF experiments, the energy-integrated cross-section $I_f$ of resonant
scattered $\gamma$-rays on the ground state, populating a state at excitation 
energy $E$ and then decaying down to a low-lying final state $f$, is given by
\begin{equation}
I_f=\frac{\pi^2\hbar^2c^2g}{E^2}\frac{\Gamma_0\Gamma_f}{\Gamma}.
\label{eq:If}
\end{equation}
Here $\Gamma_0$ and $\Gamma_f$ are the partial radiative decay widths for 
transitions from the excited state to the ground state and the final state, 
respectively, $\Gamma$ is the total radiative width of the excited state and 
$g$ is a spin-factor. In NRF experiments on even-even nuclei, the excited 
states are usually observed to decay into two states, the ground state and the 
first excited state, denoted by '1' (in deformed nuclei the first rotational 
$2^+$ state). The experimental information from NRF experiments on even-even 
nuclei consists therefore generally of the energy and the values $I_0$ and 
$I_1$ for every observed state. Often, instead of listing $I_1$ values, the 
branching ratio $R$ with
\begin{equation}
\frac{I_1}{I_0}=\frac{\Gamma_1}{\Gamma_0}=R\left(\frac{E_1}{E}\right)^3
\label{eq:I1}
\end{equation}
is given \cite{ZB90,ME95}, where $E_1$ is the energy of the $\gamma$-ray 
transition, populating the first excited state. Summing over all final states 
in Eq.\ (\ref{eq:If}), the total energy-integrated photon-absorption 
cross-section $I_{\mathrm{M1/E1,t}}=I_0+I_1+\ldots$ is obtained. 
Experimentally, one observes the partial energy-integrated photon-absorption 
cross-sections of excited states $I^{\mathrm{NRF}}_{\mathrm{M1/E1,p}}=I_0+I_1$ 
where the word 'partial' refers to the fact, that only the two decay branches 
to the ground state and to the first excited state are taken into account in 
the sum. As can be seen from Eq.\ (\ref{eq:If}),
\begin{equation}
\frac{I^{\mathrm{NRF}}_{\mathrm{M1/E1,p}}}{I_{\mathrm{M1/E1,t}}}=
\frac{\Gamma_0+\Gamma_1}{\Gamma}.
\label{eq:branch}
\end{equation}

The decay pattern in odd-nuclei is much more complex, since one can populate by
dipole radiation excited levels with, in general, three different spins. In 
NRF experiments on odd gadolinium and dysprosium isotopes \cite{ME95}, 
branching ratios have been reported to several different low-lying states. 
However, very few individual excited states show $\gamma$ decay to more than 
one of these states at the same time. In addition, the observed branching 
ratios scatter strongly. For many excited states, only the ground state 
transition has been detected. A probable explanation for this observation is 
that the dipole strength in odd nuclei is highly fragmented, thus ground state
transitions in odd nuclei are in general much weaker than in even nuclei. The 
even weaker branches to low-lying states might therefore easily fall below the 
experimental detection limit. On the other hand, due to the high fragmentation,
the chances will increase that accidentally two excited states are separated in
energy by the excitation energy of some low-lying state, and that the two 
respective peaks in the $\gamma$-ray spectrum are therefore interpreted as two 
branches from one excited state. This problem has e.g.\ already been observed
in the even nucleus $^{172}$Yb \cite{ZB90}.

In general, data on odd nuclei are very scarce (see e.g.\ \cite{EH97} and
references therein), and the results are not conclusive. The summed observed M1
strength in odd nuclei is by far weaker than in even-even nuclei. This has been
explained in terms of unresolved strength hiding in the experimental 
background. A fluctuation analysis on the background has been performed 
\cite{EH98}, and indeed the resulting summed M1 strength in $^{157}$Gd has been
found to be approximately the same value as in the neighboring even nuclei. 
However, it has not yet been shown that similar results can be achieved for the
much less fragmented M1 strength in $^{161}$Dy, which has been investigated 
with a quite similar detection threshold as $^{157}$Gd \cite{ME95}, still, a 
new analysis is in progress and might prove the opposite \cite{NE00}. In 
addition, the observed irregular branching properties of excited states in odd 
nuclei show that the NRF technique is taken to its very limit in the 
investigation of odd nuclei, and that global SC observables, like summed M1 
strengths and average excitation energies of SC states from such experiments, 
have to be taken with great care. As a last and very important point one should
mention that the M1 character of the observed radiation in NRF experiments on 
odd nuclei has never been measured experimentally. In fact, E1 and even E2 
radiation can not be excluded \cite{NS96}. In the further discussion, we will 
therefore mainly concentrate on even nuclei, where the deduced data are based 
on a more transparent experimental situation.

\section{Pygmy resonance}
\label{sec:pygmy}

\begin{figure}\centering
\includegraphics[totalheight=17.9cm]{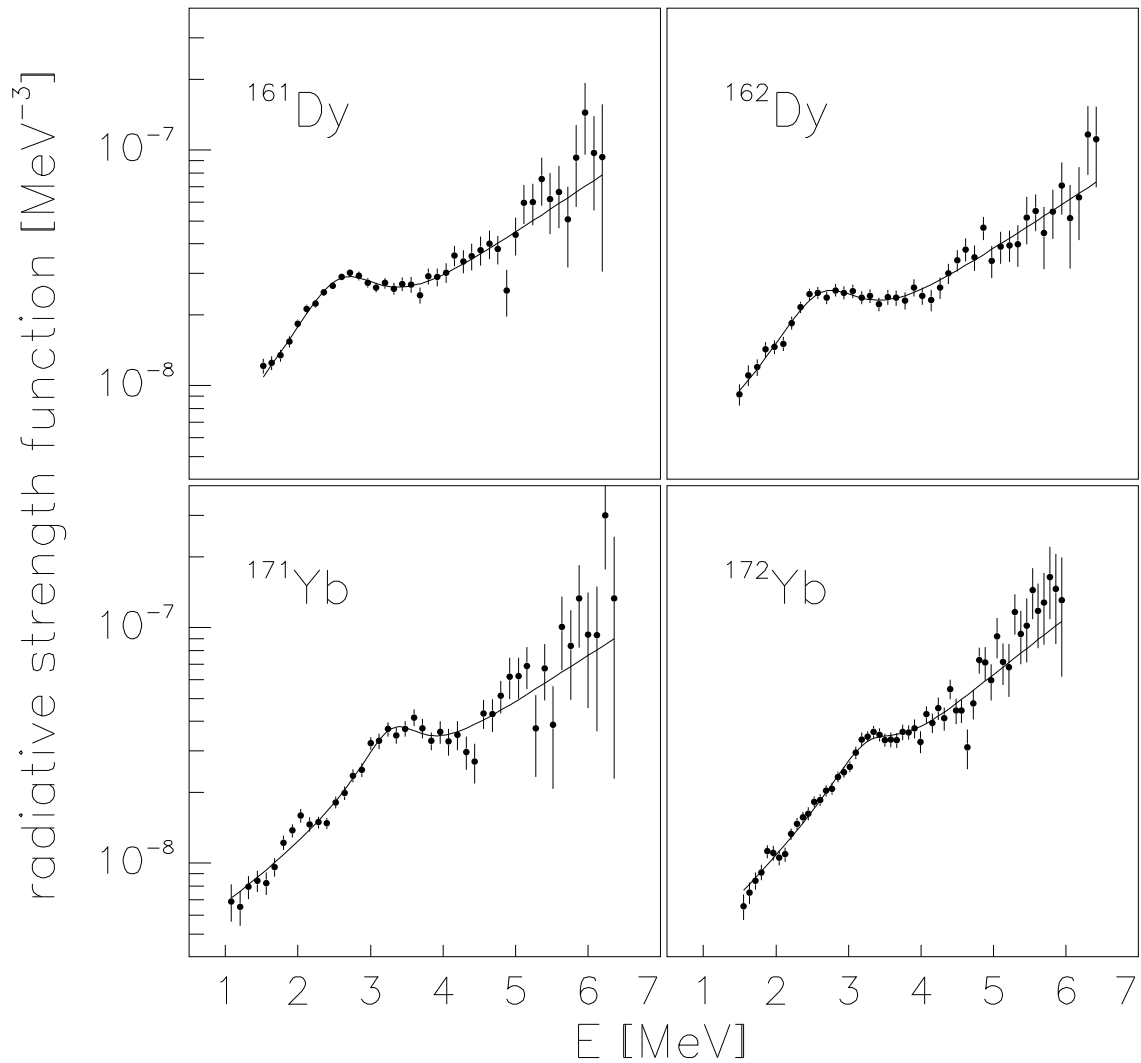}
\caption{Experimental radiative strength function data for four different rare
earth nuclei. The PY is manifested in the data at energies around 3~MeV\@. The 
solid line is a fit to the data. For details see Ref.\ [5].
}
\label{fig:data}
\end{figure}

The PY photon-absorption cross-section is usually parameterized by a Lorentzian
function $\sigma_{\mathrm{py}}(E)$
\begin{equation}
\sigma_{\mathrm{py}}(E)=\sigma_{\mathrm{py}}\left(1+
\frac{(E^2-E_{\mathrm{py}}^2)^2}{E^2\Gamma_{\mathrm{py}}^2}\right)^{-1},
\label{eq:pygmy}
\end{equation}
where $\sigma_{\mathrm{py}}$, $E_{\mathrm{py}}$ and $\Gamma_{\mathrm{py}}$ are 
the strength, centroid and width of the PY, respectively. These parameters have
been obtained by fits to experimental data \cite{VG00}, which are shown in 
Fig.\ \ref{fig:data}. Previously investigated M1 strength arises due to the 
tail of the spin-flip resonance (GMDR) \cite{KC87}. This resonance can be 
described by a Lorentzian function $\sigma_{\mathrm{sf}}(E)$ as well 
\begin{equation}
\sigma_{\mathrm{sf}}(E)=\sigma_{\mathrm{sf}}\left(1+
\frac{(E^2-E_{\mathrm{sf}}^2)^2}{E^2\Gamma_{\mathrm{sf}}^2}\right)^{-1},
\label{eq:spinflip}
\end{equation}
where $\sigma_{\mathrm{sf}}$, $E_{\mathrm{sf}}$ and $\Gamma_{\mathrm{sf}}$ are
the strength, centroid and width of the GMDR, respectively.

\begin{figure}\centering
\includegraphics[totalheight=17.9cm]{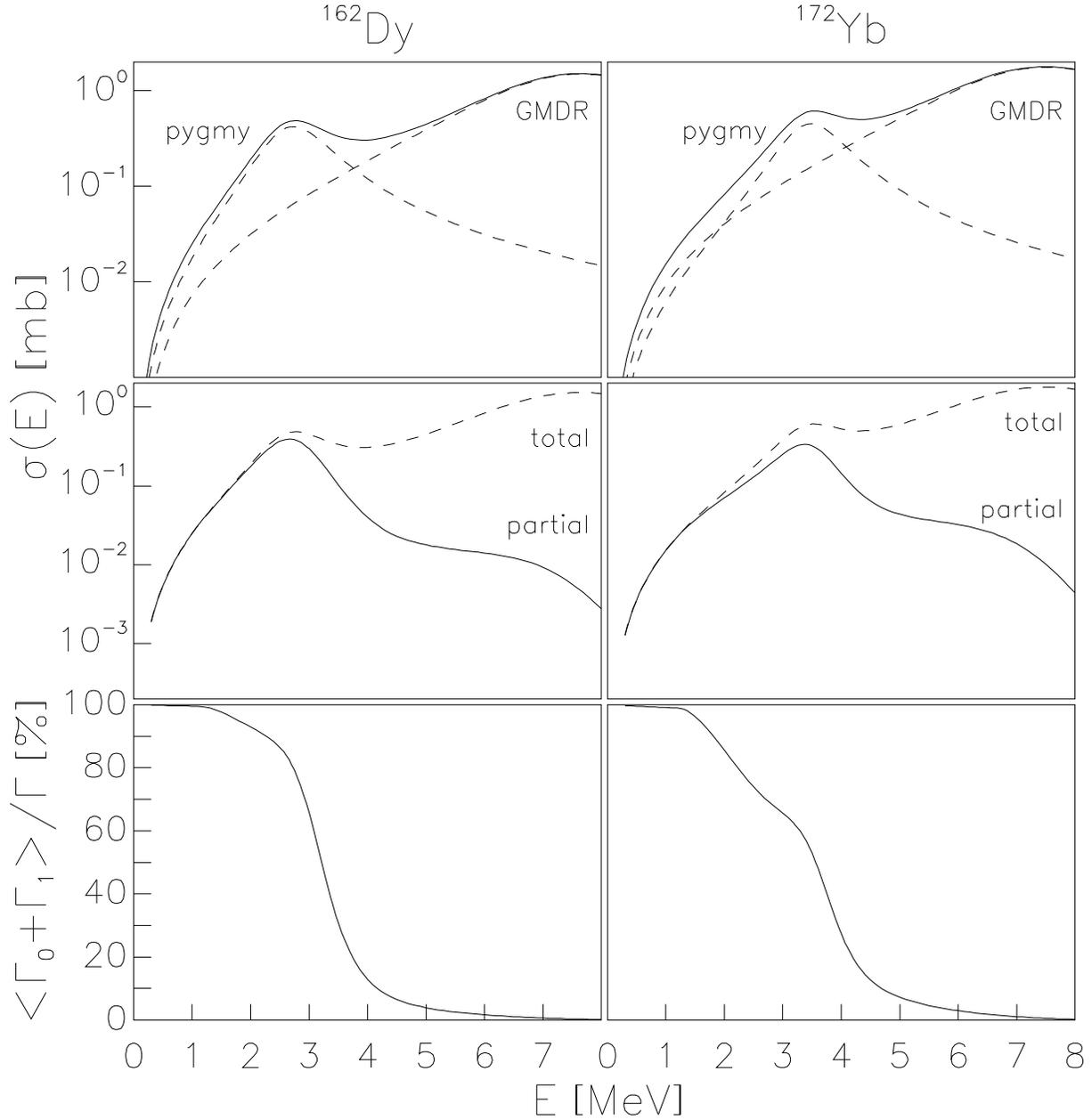}
\caption{In the upper panels, the M1 model, obtained in the quasicontinuum, is
displayed. The dashed lines indicate the Lorentzian descriptions of the PY and
the GMDR [Eq.\ (\ref{eq:pygmy}) and (\ref{eq:spinflip}), respectively], the
solid line shows the incoherent sum of both models [Eq.\ (\ref{eq:m1})]. In the
central panels, the total (dashed lines) and the partial (solid lines) M1
photon-absorption cross-sections according to Eqs.\ (\ref{eq:m1}) and
(\ref{eq:partial}) are given. In the lower panels, the ratio of these two
quantities, i.e. $\langle\Gamma_0+\Gamma_1\rangle/\Gamma$ is shown. The 
parameters needed for calculating the displayed curves are based on 
experimental data, see Ref.\ [5].
}
\label{fig:branch}
\end{figure}

Assuming that the PY is due to magnetic dipole strength, one can add the two
cross-sections incoherently and obtain the total M1 photon-absorption
cross-section $\sigma_{\mathrm{M1,t}}(E)$
\begin{equation}
\sigma_{\mathrm{M1,t}}(E)=\sigma_{\mathrm{py}}(E)+\sigma_{\mathrm{sf}}(E).
\label{eq:m1}
\end{equation}
One can, at this point, discuss if deviations from the Lorentzian description 
of the PY and interference between the PY (presumably orbital M1 strength) and 
the spin-flip resonance might occur. Neither of the two effects is, however,
manifested in the experimental data in Refs.\ \cite{JD79,IK86,MI99,VG00,MG00} 
(see e.g.\ Fig.\ \ref{fig:data}), and therefore they must be very limited in 
size and are neglected in the further calculation. In the upper panels of Fig.\
\ref{fig:branch}, the two contributions to the total M1 photon-absorption 
cross-section and their sums for the nuclei $^{162}$Dy and $^{172}$Yb are 
shown.

When comparing to the results of the NRF experiments, one should, however, in 
analogy with Eq.\ (\ref{eq:branch}) define the partial M1 photon-absorption
cross-section $\sigma^{\mathrm{QC}}_{\mathrm{M1,p}}(E)$, where branching only
to the ground state and the first excited state is taken into account, i.e.\
\begin{equation}
\sigma^{\mathrm{QC}}_{\mathrm{M1,p}}(E)=\sigma_{\mathrm{M1,t}}(E)
\frac{\Gamma_0+\Gamma_1}{\Gamma}.
\label{eq:partial}
\end{equation}
Here, the index 'QC' refers to the fact that the semi-experimental quantity 
$\sigma^{\mathrm{QC}}_{\mathrm{M1,p}}(E)$ is calculated by means of strength
functions obtained in the quasicontinuum. The total radiative width in Eq.\ 
(\ref{eq:partial}) is given by
\begin{eqnarray}
\Gamma&=&D_{1^+}\sum_{{\mathrm{X}}L}\sum_{I_f,\Pi_f}\int_0^{E}
{\mathrm{d}}E_\gamma\nonumber\\
&&E_\gamma^{2L+1}f_{{\mathrm{X}}L}(E_\gamma)\rho(E-E_\gamma,I_f,\Pi_f),
\label{eq:gt}
\end{eqnarray}
where the second sum is going over all final levels with spin and parity
$I_f,\Pi_f$ which are accessible by multipole radiation of the type X$L$ from a
$1^+$ state, $f_{{\mathrm{X}}L}$ are the respective radiative strength 
functions, $\rho$ is the level density and $D_{1^+}$ is the average spacing of
$1^+$ states. The sum of the partial radiative widths $\Gamma_0+\Gamma_1$ can 
be calculated by taking into account in Eq.\ (\ref{eq:gt}) only the two lowest 
states in the level density, i.e.\
\begin{eqnarray}
\rho(E-E_\gamma,I_f,\Pi_f)&=&\delta(E-E_\gamma)\ \delta_{I_f\Pi_f,0^+}
\nonumber\\
&&+\delta(E_1-E_\gamma)\ \delta_{I_f\Pi_f,2^+},
\label{eq:delta}
\end{eqnarray}
and performing the integral and the two sums, yielding finally
\begin{equation}
\langle\Gamma_0+\Gamma_1\rangle=D_{1^+}\left[f_{\mathrm{M1}}(E)E^3+
f_{\mathrm{M1}}(E_1)E_1^3\right].
\label{eq:g01}
\end{equation}
In Eq.\ (\ref{eq:gt}), electric quadrupole and higher-order multipole 
contributions can usually be neglected as has been done in Eq.\ (\ref{eq:g01}).

In order to be able to evaluate Eqs.\ (\ref{eq:partial}), (\ref{eq:gt}) and 
(\ref{eq:g01}), the level density and the radiative strength functions 
$f_{\mathrm{E1}}$ and $f_{\mathrm{M1}}$ have to be known. Recently, these data
have been obtained experimentally \cite{VG00} (see Fig.\ \ref{fig:data}) for 
the four nuclei $^{161,162}$Dy and $^{171,172}$Yb from the ($^3$He,$\alpha$) 
reaction by using an iteration technique. The parameters of the PY have also 
been extracted from these experiments. The quality of the data allows us to 
calculate the partial photon-absorption cross-section and make comparison to 
NRF data.

A problem encountered here is that radiative strength functions, obtained for 
$\gamma$ transitions in the quasicontinuum, have to be used at low excitation 
energies where nuclear structure effects can be predominant. For instance, the 
observed branching ratios in NRF experiments \cite{ZB90,ME95} for even nuclei
with $\Gamma_1/\Gamma_0\approx 0.5$ and 2 for $K=1$ and 0 states, respectively,
can not be reproduced accurately within this approach, where one obtains 
branching ratios of approximately 1 for both cases. The branching ratios are,
on the other hand, well estimated by the Alaga rules, describing the nucleus in
the rotational limit and thereby assuming a good $K$ quantum number. 
Unfortunately, a nuclear structure effect represented by e.g.\ a good 
$K$-quantum number can not be easily incorporated in the common definition of 
the radiative strength function. Therefore, we use here the statistical 
approach of Eq.\ (\ref{eq:g01}). Since the sum in Eq.\ (\ref{eq:g01}) is over 
two low-lying states only, we can not expect that all possible nuclear 
structure effects are averaged out by our treatment.

In Fig.\ \ref{fig:branch}, the total and partial M1 photon-absorption 
cross-sections, calculated according to Eqs.\ (\ref{eq:m1}) and 
(\ref{eq:partial}), and their ratio are shown for the nuclei $^{162}$Dy and 
$^{172}$Yb. Branching ratios further further discussed in Sect.\ 
\ref{sec:appl}. The partial energy-integrated M1 photon-absorption 
cross-section $I^{\mathrm{QC}}_{\mathrm{M1,p}}$ which should be compared to 
the experimental SC value $I^{\mathrm{NRF}}_{\mathrm{M1,p}}$, can now be 
calculated by 
\begin{equation}
I^{\mathrm{QC}}_{\mathrm{M1,p}}=\int{\mathrm{d}}E\
\sigma^{\mathrm{QC}}_{\mathrm{M1,p}}(E),
\label{eq:m1sum}
\end{equation}
where the integral covers the appropriate energy interval of interest.

\section{Comparison}
\label{sec:comp}

In the following, experimental data of the PY will be compared to the global 
observables of SC data. First, a comparison of the centroids and widths is 
performed. For the SC, one obtains an average excitation energy of 2.87~MeV for
$^{162}$Dy \cite{ME95}. This value changes by less than 10~keV when taking into
account possible additional strength where no parity assignment was possible. 
The NRF experiment on $^{172}$Yb was performed without polarimeter, thus, no 
model-independent parity assignment could be performed. If one, however, 
assumes that all states with $K=1$ are populated by M1 radiation\footnote{Here,
and in the following, we will make use of the hypothesis two in Ref.\ 
\cite{ZB90} in the disentanglement of overlapping elastic and inelastic peaks 
in the experimental NRF $\gamma$-ray spectrum. It seems that also the authors
of Ref.\ \cite{ZB90} are tending to this hypothesis (see e.g.\ their Fig.\ 5). 
Using hypothesis one would not change the conclusion drawn in this article.} 
the average excitation energy of the SC states is 3.09~MeV \cite{ZB90}. 
Including states with uncertain $K$ assignment yields a higher value of 
3.26~MeV, mainly due to the strong transition at 3.863~MeV\@. These values 
compare well with the centroids of the PY being 2.73(5)~MeV and 3.48(7)~MeV for
$^{162}$Dy and $^{172}$Yb, respectively \cite{VG00}.

The experimentally observed spreading of SC states in $^{162}$Dy and 
$^{172}$Yb, which is estimated by the energy difference between the highest and
the lowest observed SC state in the NRF experiment, are 0.67~MeV and 1.0~MeV, 
respectively \cite{ME95,ZB90}. The latter value increases to 1.3~MeV when 
taking the transition at 3.863~MeV into account. These values are in good 
agreement with the widths of the PY, which were determined to be 1.35~MeV and 
1.30~MeV for $^{162}$Dy and $^{172}$Yb, respectively \cite{VG00}.

\begin{figure}\centering
\includegraphics[totalheight=17.9cm]{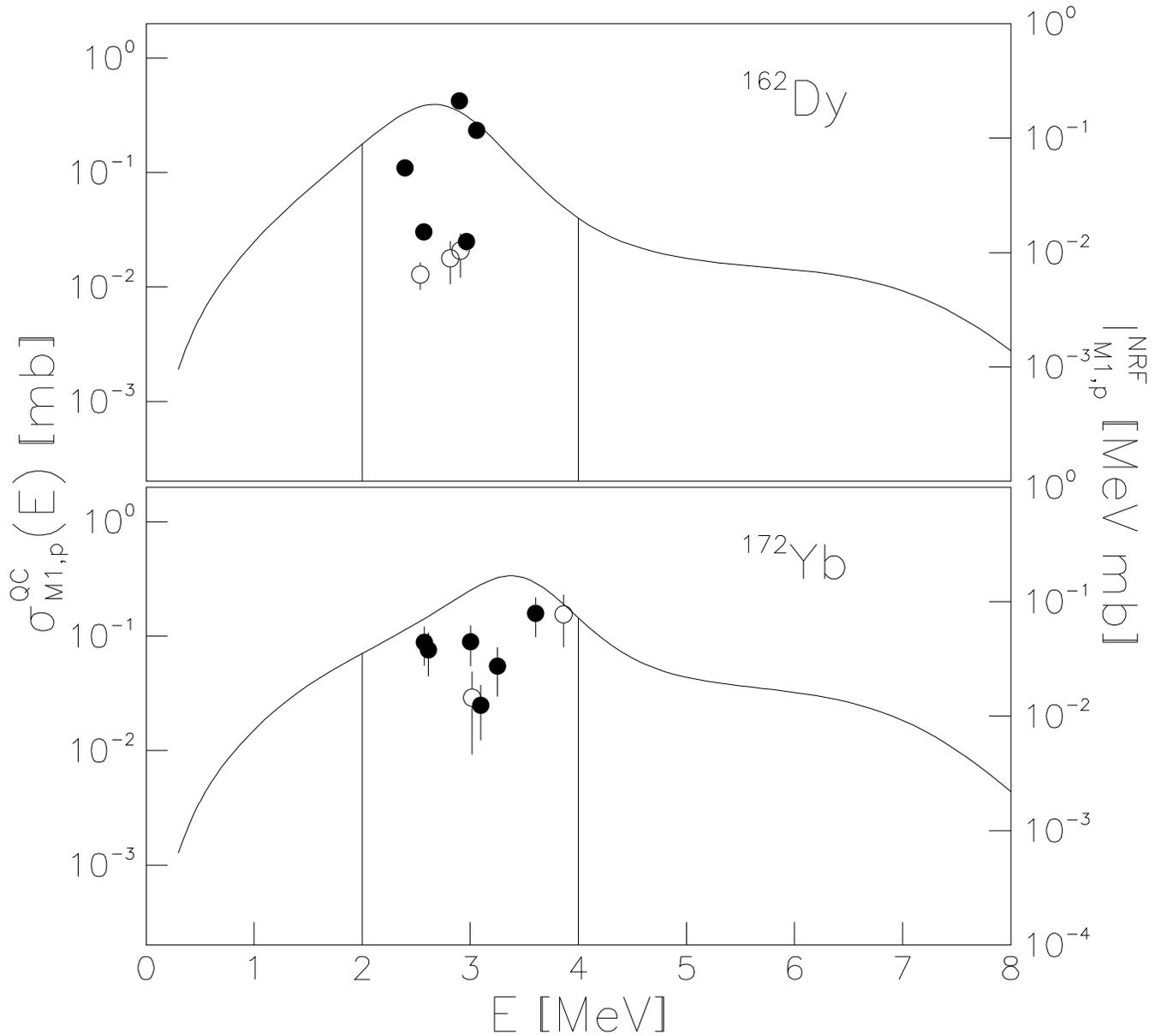}
\caption{The partial M1 photon-absorption cross-section 
$\sigma^{\mathrm{QC}}_{\mathrm{M1,p}}(E)$ according to Eq.\ 
(\ref{eq:partial}) is displayed by solid lines (scale on the left hand side).
The maxima of these curves coincide well with the energy distributions of 
partial energy-integrated photon-absorption cross-sections of SC states 
$I^{\mathrm{NRF}}_{\mathrm{M1,p}}$ [10,11] 
(scale on the right hand side). The full symbols represent states with certain
parity or $K$ assignment for $^{162}$Dy and $^{172}$Yb, respectively, the open
symbols denote states with uncertain assignments. All data necessary
for evaluating Eq.\ (\ref{eq:partial}) are taken from [5].
The vertical lines show the energy region where the integration of Eq.\
(\ref{eq:m1sum}) for the calculation of $I^{\mathrm{QC}}_{\mathrm{M1,p}}$ is
performed.}
\label{fig:dist}
\end{figure}

Figure \ref{fig:dist} shows the distribution of SC states for both nuclei. In 
the same figure, also $\sigma^{\mathrm{QC}}_{\mathrm{M1,p}}(E)$ is displayed.
The distribution of observed SC states coincides very well with the maximum and
the width of the $\sigma^{\mathrm{QC}}_{\mathrm{M1,p}}(E)$ curves.

In the second step, the strength of the PY is compared to the summed 
cross-section of SC states. For $^{162}$Dy, one obtains
$I^{\mathrm{NRF}}_{\mathrm{M1,p}}=0.41(4)\,{\mathrm{MeV\,mb}}$ for all 
transitions with M1 assignment and $0.44(5)\,{\mathrm{MeV\,mb}}$, when also 
taking into account transitions to states with undetermined parity. For 
$^{172}$Yb, these quantities read $0.25(10)\,{\mathrm{MeV\,mb}}$ and 
$0.34(15)\,{\mathrm{MeV\,mb}}$, respectively. The difference of some 
$0.09\,{\mathrm{MeV\,mb}}$ between the latter values is mostly due to the state
at 3.863~MeV with uncertain $K$ assignment and hence uncertain parity. Here, 
one should recall once more that the M1 character of transitions in $^{172}$Yb 
is not determined model independently by a polarimeter. For the PY, one has to
integrate $\sigma^{\mathrm{QC}}_{\mathrm{M1,p}}(E)$ over an appropriate energy
interval. We choose, for both nuclei, a 2~MeV wide energy region between 2~MeV 
and 4~MeV. The integration intervals are marked by vertical lines in Fig.\
\ref{fig:dist}. The choice of the intervals is motivated by the fact that in 
NRF experiments the experimental conditions allow to observe possible SC states
only in this energy region. The chosen integration region coincides also quite 
well with $E_{\mathrm{py}}\pm\Gamma_{\mathrm{py}}$, and is therefore covering 
most of the area of the PY. We obtain for $^{162}$Dy and $^{172}$Yb values of
$I^{\mathrm{QC}}_{\mathrm{M1,p}}=0.46\,{\mathrm{MeV\,mb}}$ and 
$0.42\,{\mathrm{MeV\,mb}}$, respectively. For convenience, all data are also 
collected in Table \ref{tab:result}. It is now difficult to judge the errors of
these values. Certainly, the Lorentzian shapes of the PY and the GMDR are 
somehow idealized with respect to nature. In addition, the strength functions 
used for evaluating Eq.\ (\ref{eq:partial}) are all obtained experimentally in 
the quasicontinuum, but they have to be applied to transitions to low-lying 
states in e.g.\ Eq.\ (\ref{eq:g01}). Some shortcomings of the statistical 
approach to the region of discrete levels have already been discussed in Sect.\
\ref{sec:pygmy}. Probably, a considerable systematical error is therefore 
introduced in $I^{\mathrm{QC}}_{\mathrm{M1,p}}$. Nevertheless, the difference
of 20--40\% between the calculated and the experimental values for the 
$^{172}$Yb nucleus and the agreement within 10\% for the $^{162}$Dy nucleus 
allow to hope that the systematic error of the calculation is not too high and 
within an approximate error of 30\%, we find good agreement with the SC data. 

Other reasons for discrepancy between estimated and experimental SC strength 
may exist. Firstly, some strength of the PY could in fact be E1 radiation. Such
a resonance has been suggested by P. Van Isacker et al.\ \cite{IN92} in a 
schematic calculation. Within their model, the valence neutrons oscillate with 
respect to the core of neutrons and protons. However, this excitation mode has 
not yet been observed in NRF experiments. Another possibility is that in NRF 
experiments SC states are populated from the ground state where pairing 
correlations are important. The PY strength functions are, on the other hand, 
obtained in the quasicontinuum, where depairing \cite{MB99,SB99} has been 
observed. It has been shown, that depairing yields in general a higher summed 
M1 strength and also higher excitation energies of the SC \cite{LS89}.

Unfortunately, the M1 character of the PY in rare earth nuclei could not yet be
determined experimentally. In fact, the PY in iron and lead nuclei have 
recently been measured to be of E1 character \cite{BB00,EB00}. Since the PY in
these nuclei, which have other mass numbers and deformations than rare earth 
nuclei, occur at higher excitation energies and consequently may have different
physical origins, no strong conclusions to rare earth nuclei should be drawn
from these results. An indirect proof for the M1 character of the PY in rare 
earth nuclei comes from (n,2$\gamma$) experiments on $^{162}$Dy with thermal 
neutrons, using the sum-coincidence method. Here, F. Be{\v{c}}v{\'{a}}{\v{r}} 
et al.\ \cite{BC95} have found good agreement with experiment when including a 
resonant component around 3~MeV in their M1 strength-function model. Since 
they only use schematic level density functions in their analysis, this result 
can, however, only serve as a strong hint of the M1 character of the PY in rare
earth nuclei. Still, one can conclude that different experimental observations 
commonly labeled as 'pygmy resonances' might have different physical origins.

\section{Applications}
\label{sec:appl}

\subsection{Numerical values of $\Gamma_0$ and $B({\mathrm{M}}1)$ from NRF
experiments}
\label{sub:g0bm1}

In NRF experiments, often ground-state decay widths $\Gamma_0$ or reduced
transition probabilities $B({\mathrm{M1/E1}})$ are given as results. These
quantities can be calculated from the total energy-integrated photon-absorption
cross-section $I_{\mathrm{M1/E1,t}}$ by
\begin{equation}
I_{\mathrm{M1/E1,t}}=\frac{\pi^2\hbar^2c^2g}{E^2}\Gamma_0
\label{eq:g0}
\end{equation}
and
\begin{equation}
\Gamma_0=\frac{16\pi}{9g}\left(\frac{E}{\hbar c}\right)^3B({\mathrm{M}}1),
\label{eq:bm1}
\end{equation}
where the latter Equation is only valid for M1 transitions. However, only the 
partial energy-integrated photon-absorption cross-section 
$I^{\mathrm{NRF}}_{\mathrm{M1/E1,p}}$ is measured in NRF experiments, as 
discussed in Sect.\ \ref{sec:scissors}. Usually, one therefore assumes that
$I_{\mathrm{M1/E1,t}}=I^{\mathrm{NRF}}_{\mathrm{M1/E1,p}}$ 
\cite{ZB90,ME95}, i.e.\
\begin{equation}
\Gamma_0+\Gamma_1=\Gamma
\label{eq:g0g1g}
\end{equation}
over the whole energy region where excited states are observed. On the other 
hand, Fig.\ \ref{fig:branch} shows that this assumption may become incorrect 
for energies above $\sim$3~MeV\@. Of course, the estimates in the lower panels 
of Fig.\ \ref{fig:branch} are quite rough, since experimental radiative 
strength functions obtained in the quasicontinuum are applied to the energy 
region of discrete levels. Still, the large deviation from the common 
assumption (\ref{eq:g0g1g}) can not be neglected because we have obtained good
agreement between estimated and experimental partial energy-integrated 
photon-absorption cross-sections, see Table \ref{tab:result}. As a consequence,
too low $B({\mathrm{M}}1)$ values may have been reported in NRF works. We will
therefore in the following estimate the summed $B({\mathrm{M}}1)$ value of the
SC, based on the assumption that the PY is entirely due to orbital M1 strength.

Combining Eqs.\ (\ref{eq:g0}) and (\ref{eq:bm1}) and summing over all
individual SC states, one obtains
\begin{equation}
I_{\mathrm{M1,t}}=\frac{16\pi^3}{9\hbar c}\sum E\,
B({\mathrm{M}}1)=\frac{16\pi^3}{9\hbar c}\bar{E}\sum B({\mathrm{M}}1),
\label{eq:sum}
\end{equation}
where $\bar{E}$ is the average excitation energy of the SC states. Now, we
estimate $I_{\mathrm{M1,t}}$ by integrating over the Lorentzian PY model of 
Eq.\ (\ref{eq:pygmy}) from zero to infinity, yielding
\begin{equation}
I_{\mathrm{M1,t}}=\frac{\pi}{2}\sigma_{\mathrm{py}}\Gamma_{\mathrm{py}}.
\label{eq:pysum}
\end{equation}
Assuming that the average excitation energy of the SC states $\bar{E}$ is equal
to the centroid $E_{\mathrm{py}}$ of the PY, we finally obtain an estimate for 
the summed $B({\mathrm{M}}1)$ value of SC states
\begin{equation}
\sum B({\mathrm{M}}1)=\frac{9\hbar c}{32\pi^2}
\frac{\sigma_{\mathrm{py}}\Gamma_{\mathrm{py}}}{E_{\mathrm{py}}}.
\label{eq:bm1sum}
\end{equation}
Using experimental PY parameters from Ref.\ \cite{VG00}, summed 
$B({\mathrm{M}}1)$ strengths of $7\pm2\,\mu_{\mathrm{N}}^2$ and 
$6\pm2\,\mu_{\mathrm{N}}^2$ for $^{162}$Dy and $^{172}$Yb, respectively, are 
found. Here, one should stress that these values follow rather directly from 
experiment, since the PY parameters are obtained by a fit to normalized, purely
experimental strength function data. Neither branching ratios are needed, nor 
the assumption of a Lorentzian shape of the PY does influence this result much.
Although these data are rather high compared to recent estimates 
\cite{PB98,EK99} using effective relative $g$-factors of $\sim 0.7$, 
theoretical values around e.g.\ 6--7$\,\mu_{\mathrm{N}}^2$ have indeed been 
discussed in the literature for dysprosium and ytterbium nuclei in a sum-rule 
approach \cite{LS89}, assuming free values of the $g$-factors. If our 
assumption for calculating the summed SC strength holds, the reason for 
observing summed strengths around $3\,\mu_{\mathrm{N}}^2$ in NRF experiments 
would not be the manifestation of effective $g$ factors, but rather partly the 
lack of detected branching from SC states to higher lying states than the first
excited state and partly the fact that some strength lies outside the 
investigated energy region of 2-4~MeV\@. 

\subsection{E1 strength in NRF experiments on even-even nuclei}
\label{sub:e1}

The assumption of a dominant M1 character of the PY can also be confirmed by a 
reproduction of the observed E1 strength in NRF experiments using the strength 
functions from the quasicontinuum. For this reason, we calculate the partial E1
photon-absorption cross-section $\sigma^{\mathrm{QC}}_{\mathrm{E1,p}}(E)$ 
according to the model of S.G. Kadmenski{\u{\i}} et al.\ \cite{KM83} 
\begin{equation}
\sigma^{\mathrm{QC}}_{\mathrm{E1,p}}(E)=\sum_{i=1,2}
\frac{0.7\sigma_{i,{\mathrm{E1}}}\Gamma_{i,{\mathrm{E1}}}^2(E^2+4\pi^2T^2)}
{EE_{i,{\mathrm{E1}}}(E^2-E_{i,{\mathrm{E1}}}^2)^2}
\frac{\langle\Gamma_0+\Gamma_1\rangle}{\Gamma}
\label{eq:kmfp}
\end{equation}
which describes well the tail of the giant electric dipole resonance and 
reproduces the non-vanishing E1 strength at low energies, as observed by Yu.P. 
Popov \cite{Po82} in (n,$\gamma\alpha$) reactions. In Eq.\ (\ref{eq:kmfp}),
$\sigma_{i,{\mathrm{E1}}}$, $\Gamma_{i,{\mathrm{E1}}}$ and 
$E_{i,{\mathrm{E1}}}$ are the cross-sections, the widths and the centroids of 
the two parts of the giant electric dipole resonance, respectively, and $T$ is 
the nuclear temperature. In the following, we use a constant value for the 
nuclear temperature, which has been obtained by a fit to the experimental data 
of Fig.\ \ref{fig:data} \cite{VG00}. Similar approaches have also been very 
successful in describing $\gamma$-ray spectra of radiative neutron-capture 
experiments and in the calculation of isomeric cross-sections \cite{Gr99,Gr00}.
Integrating Eq.\ (\ref{eq:kmfp}) over an appropriate energy interval should now
yield a value approximately equal to the sum of all partial energy-integrated 
E1 photon-absorption cross-sections $I^{\mathrm{NRF}}_{\mathrm{E1,p}}$ observed
in NRF experiments and defined in Eq.\ (\ref{eq:branch}).

\begin{figure}\centering
\includegraphics[totalheight=17.9cm]{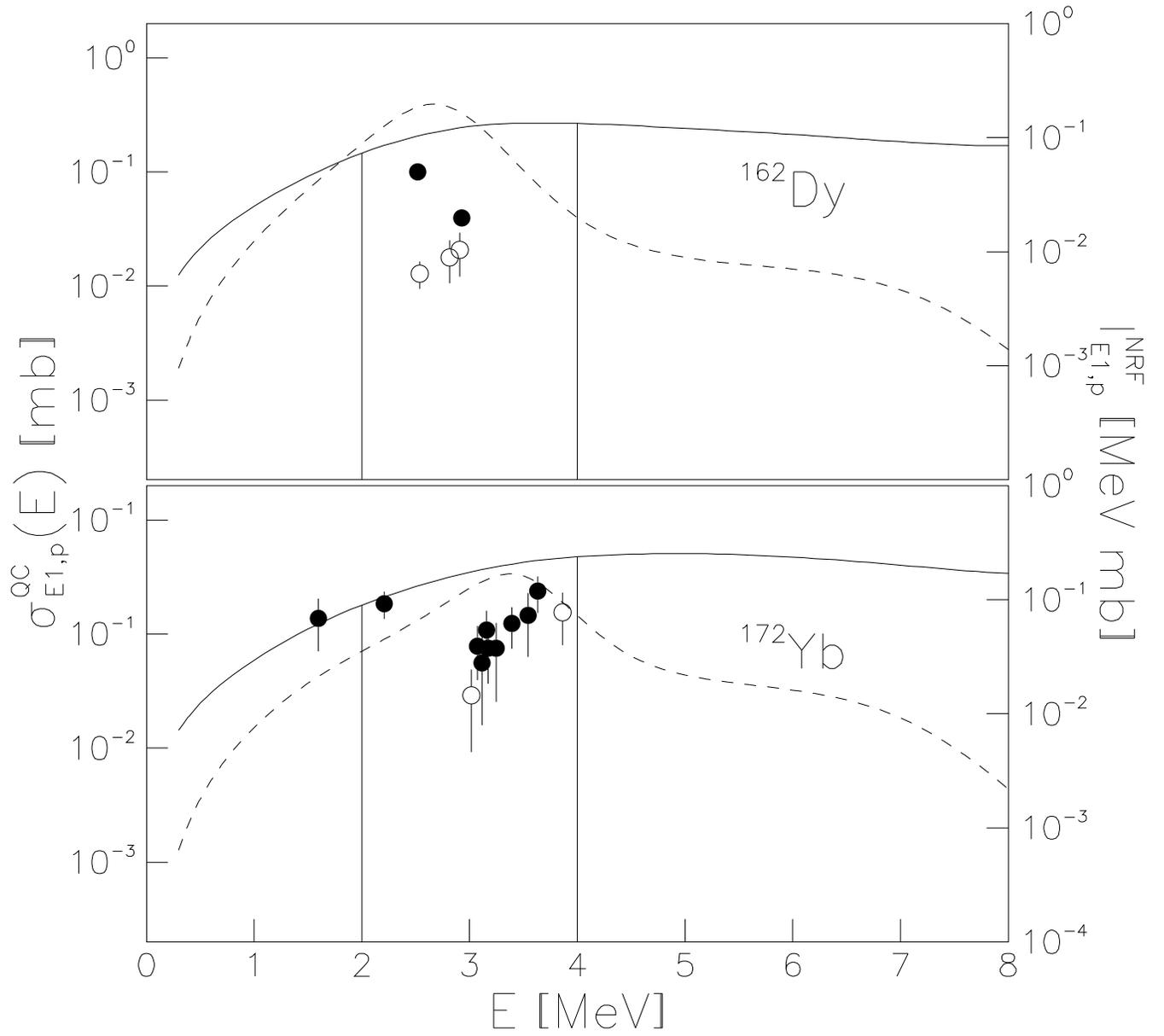}
\caption{The same as in Fig.\ \ref{fig:dist} but for E1 radiation. The dashed 
lines denote the $\sigma^{\mathrm{QC}}_{\mathrm{M1,p}}(E)$ curves for 
comparison.}
\label{fig:e1m1even}
\end{figure}

In Fig.\ \ref{fig:e1m1even} the distribution of partial energy-integrated E1
photon-absorption cross-sections is displayed. From the NRF experiments, we 
obtain $I^{\mathrm{NRF}}_{\mathrm{E1,p}}=0.07(1){\mathrm{MeV\,mb}}$ and
$0.6(3){\mathrm{MeV\,mb}}$ for $^{162}$Dy and $^{172}$Yb, respectively, when 
taking into account only levels with certain parity assignments in the case of 
$^{162}$Dy and certain $K$ assignments in the case of $^{172}$Yb. These values 
read $0.10(2){\mathrm{MeV\,mb}}$ and $0.7(3){\mathrm{MeV\,mb}}$ for the two 
nuclei when also taking into account levels with uncertain assignments. 
Integrating Eq.\ (\ref{eq:kmfp}) from 2~MeV to 4~MeV, as it is done in the case
of M1 radiation in Sect.\ \ref{sec:comp}, we obtain from the quasicontinuum 
$0.5\,{\mathrm{MeV\,mb}}$ and $0.7\,{\mathrm{MeV\,mb}}$ for $^{162}$Dy and 
$^{172}$Yb, respectively. Also here, we judge the errors of these quantities to
be around 30\% due to systematical errors, see Sect.\ \ref{sec:comp}. For 
easier comparison all values can also be found in Table \ref{tab:result}.

Clearly, the observed strength in NRF experiments on $^{172}$Yb is very well 
described by the strength functions from the quasicontinuum, whereas this is 
not at all true for $^{162}$Dy. We have no good explanation for the discrepancy
in the latter nucleus. One can of course argue that the E1 strength function 
from the quasicontinuum might not describe well the observed E1 strength in the
region of discrete levels. We find, however, that our approach works well for 
$^{172}$Yb. Certainly, the situation would be even worse when assuming a 
substantial part of the PY to be of E1 character. The good agreement in terms 
of the M1 strength in both nuclei and the E1 strength in $^{172}$Yb would be 
degraded, while the overestimation of observed E1 strength in NRF experiments 
on $^{162}$Dy by the strength function from the quasicontinuum would even 
increase. Another possibility is that a substantial part of the E1 strength in 
the NRF experiment is hidden in the experimental background. If we, however, 
accept a detection limit of some $0.005\,{\mathrm{MeV\,mb}}$, also this seems 
quite unrealistic. At last, one can imagine some distinct nuclear structure 
effect responsible for hindering the population of $1^-$ states at around 3~MeV
from the ground state. The physical nature of this possible effect is, however,
not clear at all.

\subsection{Missing strength in NRF experiments on odd nuclei}
\label{sub:odd}

An important issue of the SC is the question of missing strength in odd nuclei
observed in NRF experiments. The solution of this problem is that in the case
of odd nuclei, indeed a large portion of M1 strength is hidden in the
experimental background \cite{EH97,EH98}. As a consequence, the summed M1
strength in odd nuclei might be comparable to the observed strength in
neighboring even nuclei. The method for recovering the full strength yields,
however, large errors and gives at the present unsatisfactory results for 
$^{161}$Dy \cite{EH97}. The latter statement might be disproved be a new 
experimental analysis \cite{NE00}. The PY parameters, on the other hand, show 
no odd-even effect at all \cite{VG00} and the summed $B({\mathrm{M}}1)$ 
strengths in odd nuclei, calculated according to Eq.\ (\ref{eq:bm1sum}) are 
$9\pm2\,\mu_{\mathrm{N}}^2$ and $7\pm2\,\mu_{\mathrm{N}}^2$ for $^{161}$Dy and 
$^{171}$Yb, respectively, which are equal within the errors to those of 
neighboring even nuclei. In the following, a possible explanation of the 
observations in NRF experiments on odd nuclei is given by means of strength 
functions from the quasicontinuum. 

\begin{figure}\centering
\includegraphics[totalheight=17.9cm]{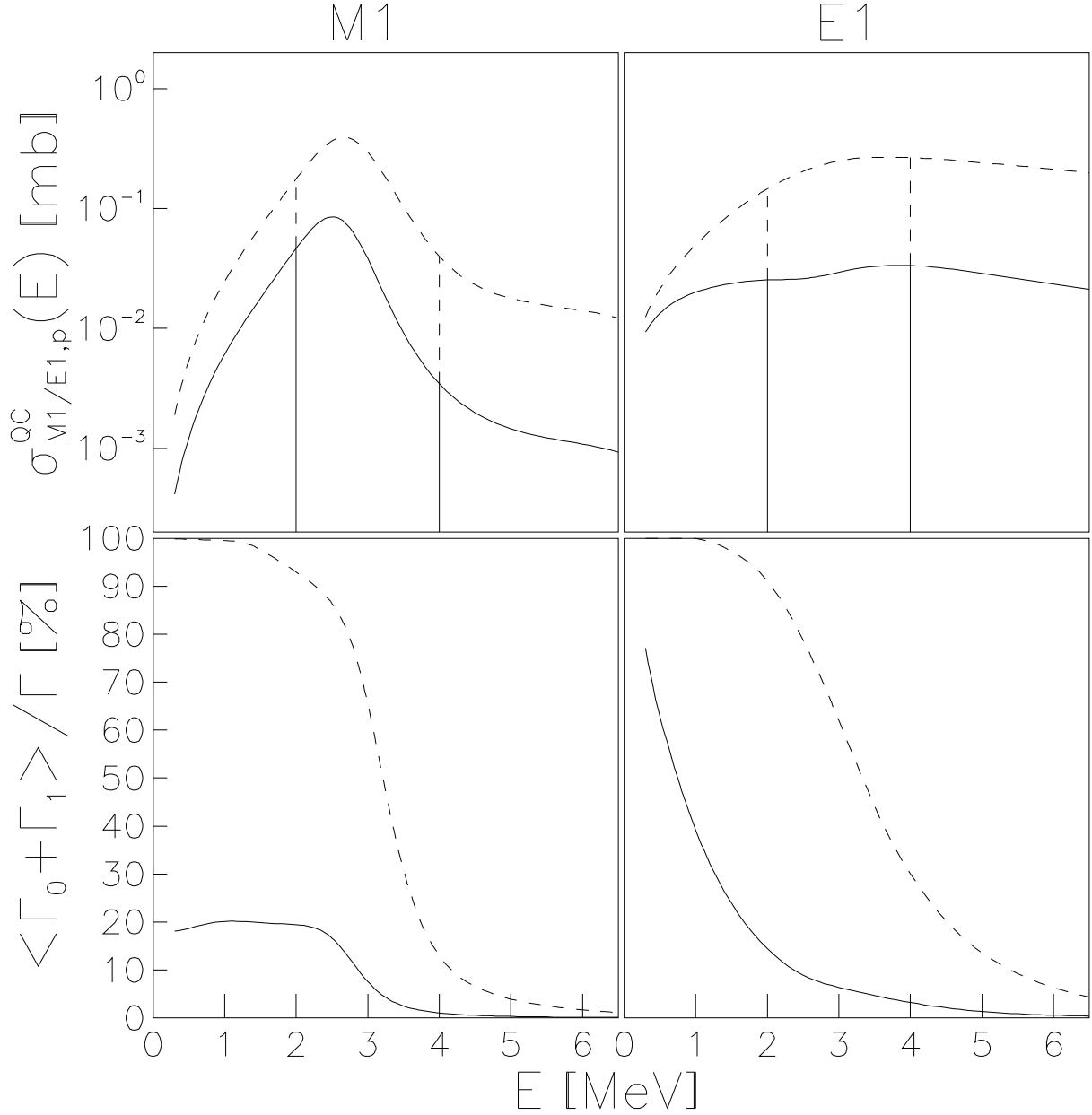}
\caption{Upper panels: The partial M1 and E1 photon-absorption cross-sections
$\sigma^{\mathrm{QC}}_{\mathrm{M1,p}}(E)$ and 
$\sigma^{\mathrm{QC}}_{\mathrm{E1,p}}(E)$ according to Eqs.\ 
(\ref{eq:partial}) and (\ref{eq:kmfp}), respectively. Lower panels: The 
branching ratio $\langle\Gamma_0+\Gamma_1\rangle/\Gamma$. The solid lines 
denote $^{161}$Dy data and the dashed lines $^{162}$Dy data. The parameters 
needed for calculating the displayed curves are based on experimental data, see
Ref.\ [5].
}
\label{fig:odd}
\end{figure}

The calculated partial M1 and E1 photon-absorption cross-sections using Eqs.\ 
(\ref{eq:partial}) and (\ref{eq:kmfp}) are displayed in Fig.\ \ref{fig:odd}. 
Also, the sum of all energy-integrated partial photon-absorption cross-sections
from Ref.\ \cite{ME95} is calculated, thereby taking into account all observed 
branching. Integrating the E1 and M1 curves of Fig.\ \ref{fig:odd} as usual 
from 2~MeV to 4~MeV yields $0.062\,{\mathrm{MeV\,mb}}$ and 
$0.083\,{\mathrm{MeV\,mb}}$, respectively, and again a systematical error of 
30\% is assumed. The sum of both contributions is $0.145\,{\mathrm{MeV\,mb}}$ 
and agrees well with the value from the NRF experiment 
$0.11(2){\mathrm{MeV\,mb}}$. Unfortunately, in NRF experiments on odd nuclei, 
neither parity nor $K$ assignments can be performed reliably \cite{ME95}. Thus 
one can not distinguish between M1 and E1 strength. Again, the numerical values
are collected in Table \ref{tab:result} for convenience.

Figure \ref{fig:odd} also shows the branching ratio
$\langle\Gamma_0+\Gamma_1\rangle/\Gamma$ in $^{161}$Dy. This branching ratio is
generally much smaller in odd nuclei than in even nuclei $^{162}$Dy due to the 
higher level density in odd nuclei entering Eq.\ (\ref{eq:gt}). Here, the 
physical reason for the missing strength in odd nuclei becomes obvious. We find
that the missing strength in NRF experiments might not only be due to 
unobserved elastic photon scattering below the experimental threshold, as 
suggested in Refs.\ \cite{EH97,EH98}, but also due to unobserved branching to 
higher lying states than the first excited states; a possibility which was not 
mentioned in earlier works and which can contribute much more to the summed 
strength than in even nuclei. These branches could, in principle, be detected 
in a similar type of fluctuation analysis on the experimental background as in 
Ref.\ \cite{EH97}; still, one has to expect a general tendency of these 
branches to appear at lower $\gamma$-ray energies in the $\gamma$-ray spectrum.

Additional sources of uncertainty in the calculation of partial 
photon-absorption cross-sections in odd nuclei comes from the use of the 
branching ratio $\langle\Gamma_0+\Gamma_1\rangle/\Gamma$ and the fact that only
7/2 states are taken into account as intermediate states\footnote{Since the 
ground state spin of $^{161}$Dy is 5/2, one can also populate 3/2 and 5/2 
states by absorption of dipole radiation.}. As discussed in Sect.\ 
\ref{sec:scissors}, observed branching for odd nuclei in NRF experiments is 
much more complex than for even nuclei since it is dependent of the spin of the
intermediate state and it does not only involve decay to the first excited 
state. The use of $\langle\Gamma_0+\Gamma_1\rangle/\Gamma$ calculated for
intermediate 7/2 states should therefore be regarded as an approximation to the
average branching properties of all observed states in odd nuclei. 

\section{Conclusion}

In conclusion, global parameters of the SC have been compared to PY parameters.
A good agreement has been found for the centroids, the widths and the partial
energy-integrated M1 photon-absorption cross-sections. Therefore, although the
M1 character of the PY has not been verified experimentally, we conclude, that 
the PY in rare earth nuclei with high probability originates from orbital M1 
strength. Still, direct measurements of the M1 character of the PY by e.g.\ the
sum-coincidence method, using experimental level densities, is very desirable.

Further, we have shown that branching to highly excited states from SC states
might be important in the estimation of B(M1) strengths. Our approach, which 
does not depend on the knowledge of branching ratios, gives for even and odd 
nuclei summed $B({\mathrm{M}}1)$ strengths of 6-9$\mu_{\mathrm{N}}^2$ for the 
SC in the quasicontinuum, which follow rather directly from experimental data 
and are in good agreement with a sum-rule approach using free $g$-factors. The 
E1 strength observed in NRF experiments could also be reproduced by the 
strength functions from the quasicontinuum in the case of $^{172}$Yb, whereas 
in the case of $^{162}$Dy a discrepancy has been observed. 

Also the missing strength in NRF experiments on odd nuclei compared to even 
nuclei has been interpreted in terms of additional branching to higher lying 
states than the first excited states. Since branching might play a crucial role
in the interpretation of NRF data, we would like to confront our branching 
estimates to experimental measurements of branching ratios from the 
sum-coincidence method. Here, one should especially keep in mind that the 
radiative strength functions used for estimating branching ratios in this work 
are only based on experimental data down to energies of $\sim 1.5$~MeV, whereas
below, extrapolations of the experimental data by theoretical models have to be
used.

The investigation of the SC in the quasicontinuum is complementary to the
conventional NRF method. It might yield valuable data on branching ratios, 
summed M1 strengths and possible odd-even effects, as well as the influence of 
depairing on the summed M1 strength.

\acknowledgements

The authors are grateful to J. Kopecky, E. Lipparini, P. von Neumann-Cosel and 
J. Enders for interesting discussions. This work is supported by the Norwegian 
Research Council (NFR).

\end{multicols}

\begin{table}
\caption{Partial energy-integrated photon-absorption cross-sections for M1 and 
E1 radiation from NRF experiments and estimated from strength functions 
(details see text). All values are in MeV~mb. The errors of the calculated 
values are estimated to be around 30\%.}
\begin{tabular}{cccccccc}
&&\multicolumn{2}{c}{$^{162}$Dy}&\multicolumn{2}{c}{$^{172}$Yb}&\multicolumn{2}{c}{$^{161}$Dy}\\
&&M1&E1&M1&E1&M1&E1\\\hline
\rule{0mm}{9pt}$I^{\mathrm{NRF}}_{\mathrm{M1/E1,p}}$&only good assignment&0.41(4)&0.07(1)&0.25(10)&0.6(3)&&\\
&also uncertain assignment&0.44(5)&0.10(2)&0.34(15)&0.7(3)&\multicolumn{2}{c}{0.11(2)}\\
$I^{\mathrm{QC}}_{\mathrm{M1/E1,p}}$&&0.46&0.5&0.42&0.7&0.083&0.062\\
\end{tabular}
\label{tab:result}
\end{table}

\clearpage

\clearpage

\clearpage

\clearpage

\clearpage

\end{document}